\documentclass[preprint,prc,showpacs,preprintnumbers,superscriptaddress]{revtex4-1}
\usepackage{graphicx,latexsym,amssymb,amsmath,color,multirow,mathrsfs,booktabs,ifpdf}
\usepackage{dcolumn}
\usepackage{bm}
\usepackage{longtable}

\def\oc{$^{16}$O+$^{12}$C\ }
\def\ac{$\alpha+^{12}$C\ }

\def\cc{$^{12}$C+$^{12}$C\ }
\def\oo{$^{16}$O+$^{16}$O\ }
\def\AA{nucleus-nucleus\ }

\begin{document}
\title{Elastic transfer and parity dependence of the \AA optical potential}
\author{Nguyen Tri Toan Phuc} \email{nguyentritoanphuc@yahoo.com} 
\affiliation{Department of Nuclear Physics, University of Science, 
VNU-HCM, 227 Nguyen Van Cu, District 5, 700000 Ho Chi Minh City, Vietnam}
\affiliation{Institute for Nuclear Science and Technology, VINATOM \\
 179 Hoang Quoc Viet, Cau Giay, 100000 Hanoi, Vietnam}
\author{R.S. Mackintosh} 
\affiliation{School of Physical Sciences, The Open University, Milton Keynes, 
 MK7 6AA, United Kingdom}
\author{Nguyen Hoang Phuc} 
\author{Dao T. Khoa}
\affiliation{Institute for Nuclear Science and Technology, VINATOM \\
179 Hoang Quoc Viet, Cau Giay, 100000 Hanoi, Vietnam}

\begin{abstract}
\begin{description}
\item[Background] A recent coupled reaction channel (CRC) study shows that the 
enhanced oscillation of the elastic \oc	cross section at backward angles is due 
mainly to the elastic $\alpha$ transfer or the core exchange. Such a process 
gives rise to a parity-dependent term in the total elastic $S$-matrix, an indication
of the parity dependence of the \oc optical potential (OP). 
\item[Purpose] To explicitly determine the core exchange potential (CEP) induced by 
the symmetric exchange of the two $^{12}$C cores in the elastic \oc scattering 
at $E_{\rm lab}= 132$ and 300 MeV, and explore its parity dependence.
\item[Method] $S$-matrix generated by CRC description of the elastic \oc scattering 
is used as the input for the inversion calculation to obtain the effective local 
OP that contains both the Wigner and Majorana terms.
\item[Results] The high-precision inversion results show a strong contribution 
by the complex Majorana term in the total OP of the \oc system, and thus provide for 
the first time a direct estimation of the parity-dependent CEP.  
\item[Conclusions] The elastic $\alpha$ transfer or exchange of the two 
 $^{12}$C cores in the \oc system gives rise to a complex parity dependence 
 of the total OP. This should be a general feature of the OP for the light 
 heavy-ion systems that contain two identical cores. 
\end{description}

\end{abstract}
\date{\today}
\maketitle

\section{Introduction} \label{intro.sec}
It is well established that in the elastic scattering of light heavy-ion (HI) systems 
where the projectile and target differ only by one nucleon or a nuclear cluster, elastic nucleon or cluster transfer processes can take place between the two identical cores \cite{vOe75}. Such pairs of nearly identical nuclei are known as ``core-identical'' nuclei.
The widely used approach to describe the elastic transfer is to add coherently, 
in the distorted wave Born approximation (DWBA), the elastic transfer amplitude 
to that of the elastic scattering. The interference between these two amplitudes 
gives a rapidly oscillating cross section as observed in both the excitation function 
and elastic scattering cross section at backward angles. A more consistent approach 
is to study the elastic transfer within the coupled reaction channels (CRC) 
formalism \cite{Sat83,Tho88,Tho09}, which provides the most accurate physics 
description and a clear insight into this process \cite{Phu18}.
 
At low energies, the enhanced oscillating cross section at backward angles known as anomalous large angle scattering (ALAS) has been observed in the elastic scattering of various light HI systems \cite{Brau82}. The elastic transfer is the main physical origin of the ALAS observed in the elastic scattering of core-identical systems such as \oc at low energies \cite{vOe75,Phu18}. Such transfer processes can be found not only in the elastic scattering but also in inelastic scattering \cite{vOe75,Ima87} and fusion \cite{Chr95,Row15}. 
The ALAS pattern can be reproduced in the optical model (OM) calculation using an explicitly parity-dependent optical potential (OP) \cite{vOe75,Brau82}. Such a procedure was studied by Frahn \cite{Frahn84} and results in a modified elastic $S$-matrix that contains a parity-dependent component. 
The elastic transfer (or the core exchange) reaction has been used to study the cluster- or nucleon 
spectroscopic factors \cite{vOe75}, molecular orbitals \cite{Ima87,vOe06}, pairing effect 
\cite{vOe01,Erma16}, and cluster correlations \cite{Szi99,Phu18} in stable and exotic nuclei 
\cite{Kee09}. Given its peripheral nature, HI-induced elastic transfer process can also be 
used to extract the asymptotic normalization coefficient, an important ingredient 
in the nuclear astrophysics studies \cite{Muk97,How13}. The relation between ALAS, elastic transfer, 
and parity dependence of the OP has been shown in a recent study of \ac scattering, where the 
elastic transfer is used to study the Hoyle state \cite{Bel10} while a parity-dependent potential 
is required to reproduce the important $^{12}$C($\alpha$,$\gamma$)$^{16}$O radiative 
capture process \cite{Kat08}. Last but not least, the observation of nuclear rainbow scattering 
in the core-identical systems like \oc or $^{13}$C+$^{12}$C \cite{Bra97,Kho07r} requires 
a better understanding of the low-energy elastic transfer that deteriorates the rainbow 
pattern at large angles \cite{Phu18}, and its link to a parity-dependent OP in the OM 
description of elastic scattering data.

In a conventional single-channel OM calculation, it has been suggested that the elastic 
transfer process generates an additional term in the total OP \cite{vOe70,vOe73,Fra80,Ful72}, 
which we refer hereafter to as the core exchange potential (CEP). The CEP originates 
from the exchange of the two identical cores and should be, therefore, parity-dependent 
(i.e., containing a Majorana term). Moreover, it has been suggested in Refs.~\cite{vOe73,Ful72} 
that the CEP is also complex. The existence of a parity-dependent potential due to the core 
exchange was already pointed out in the early studies using the microscopic RGM
and GCM methods \cite{Lem79,Bay77,Bay86,Aoki83} that treat exactly the antisymmetrization 
implied by the Pauli principle. In general, the exchange of nucleons in a microscopic model 
or identical (structureless) cores in a macroscopic model leads readily to the parity 
dependence of the OP. Nevertheless, the explicit derivation of the CEP within the general 
Feshbach formalism \cite{Ful72,Fra80} that is capable of reproducing the scattering 
data still remains a challenge. Given the description of the core exchange by different 
models of elastic transfer \cite{vOe70,vOe73,vOe75,Ful72,Fra80}, many OM analyses 
of elastic scattering data measured for the core-identical systems at low energies 
were done using the real CEP based either on the linear combination of nuclear orbitals 
(LCNO) \cite{Bar86,Fer90,Gao92} or phenomenological parity-dependent potentials 
\cite{Kon85,Vit86}. Although the use of these parity-dependent potentials drastically 
reduces the complexity of calculations, their connections to the underlying core exchange 
process is still not yet fully understood. So far, the CEP has never been directly 
derived from the elastic transfer calculation using DWBA or CRC methods, and a better 
understanding of the physics origin of the CEP in an elastic scattering process is 
of high interest.  

The iterative-perturbative (IP) inversion of the scattering $S$-matrix to the 
equivalent local OP has been proven to be accurate, especially, when applied 
to the $S$-matrix given by CRC calculations \cite{IPreview}. Therefore, it is 
of high interest to use this inversion method to explicitly determine the CEP 
in the OP of a typical core-identical system. A recent extensive CRC study of the 
elastic \oc scattering with up to ten reaction channels included \cite{Phu18} has 
clearly demonstrated a strong impact of the elastic $\alpha$ transfer on the elastic 
scattering cross section at different energies. In the present work, we apply the IP 
inversion method to the complex $S$-matrix given by the CRC calculation of elastic
\oc scattering at $E_{\rm lab}=132$ and 300 MeV \cite{Phu18} to deduce the radial 
strength of the local CEP that is directly generated by the elastic $\alpha$ 
transfer, and explore the parity dependence of the OP. 

\section{Core-core symmetry and parity dependence of the OP} 
\label{sec2}
\begin{figure}[bht]\vspace*{0cm}
	\includegraphics[width=0.82\textwidth]{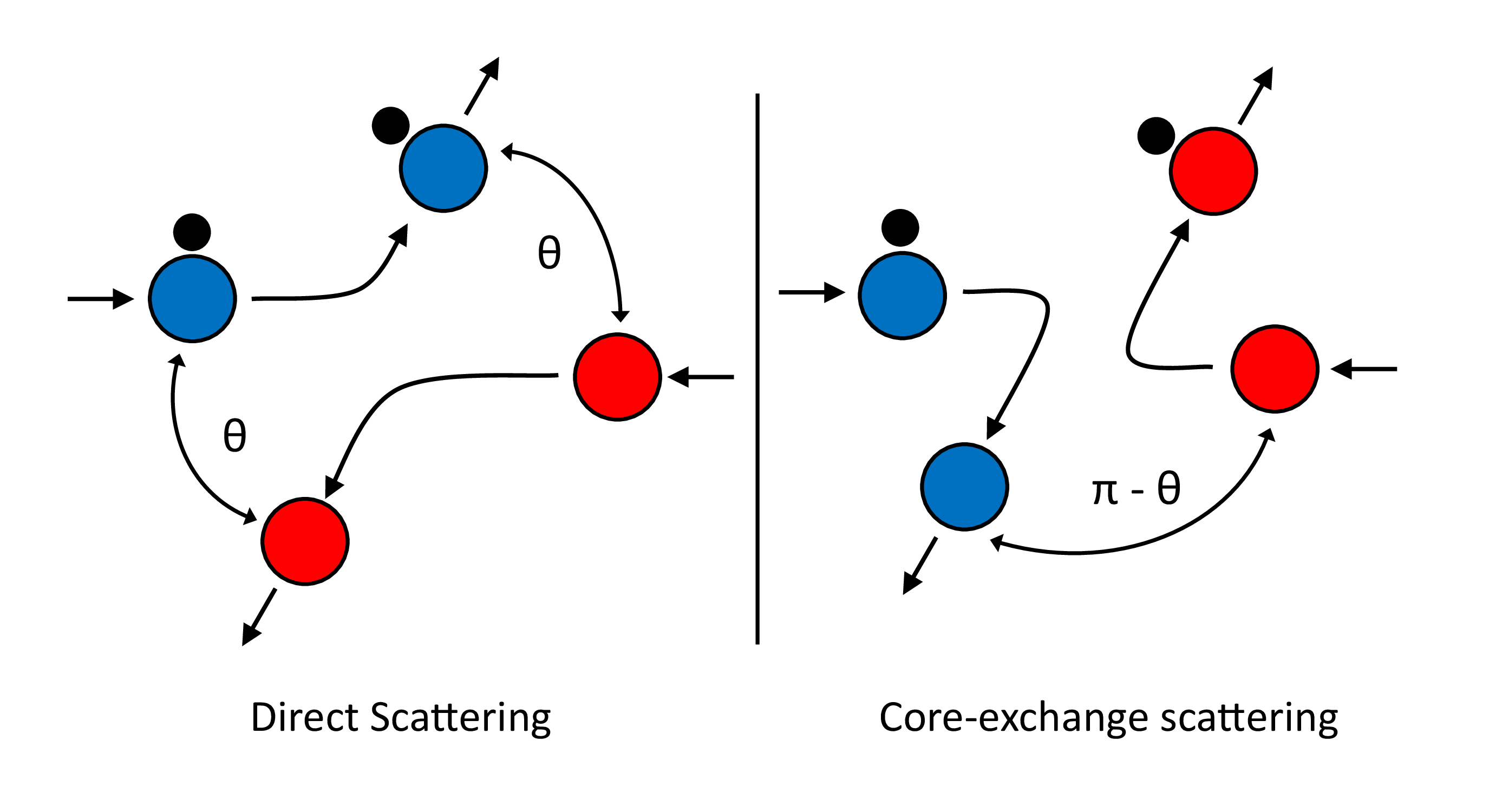}\vspace*{0cm}
	\caption{Elastic transfer viewed as the exchange of the two identical cores. 
 The red and blue spheres are the two cores in different initial states, and 
 the black circle represents either the valence nucleon or cluster being 
 transferred.} \label{f1}
\end{figure}

We show here briefly that the parity dependence caused by the elastic transfer 
process is a natural consequence of the core-core symmetry that shows up in the 
exchange of the two identical cores. In the center-of-mass (c.m.) frame of a  
core-identical system like \oc, the elastic transfer of the valence nucleon or 
cluster is equivalent to the exchange of the two cores as illustrated in 
Fig.~\ref{f1}. Quantum mechanically, such an exchange process is possible by acting 
the core-exchange Majorana operator $P_{\rm c}$ on the scattering wave function, 
in a manner similar to the projectile-target exchange in the elastic scattering 
of two identical nuclei.      
For this purpose, we consider the transition amplitude of the \AA scattering 
in the following general form \cite{Sat83,Tho09}    
\begin{equation}
 T=\langle \phi|V|\Psi_L\rangle, \label{eq1}
\end{equation} 
where $\Psi$ and $\phi$ are the total scattering wave function and that of the 
entrance, respectively. Orbital momentum $L$ of each partial wave is specified
explicitly in Eq.~(\ref{eq1}) to trace the parity of the scattering wave function 
and the bra-ket notation is assumed, therefore, to contain also the summation over 
all partial waves. When both cores have the same nonzero spin $I_{\rm c}$, the total 
wave function must be properly symmetrized to account for both the direct elastic 
scattering and exchange of the two identical cores 
\begin{equation}
\Psi_L(\bm{R})\to \Psi_L(\bm{R}) + (-1)^{2I_{\rm c}}P_{\rm c}\Psi_L(\bm{R}), 
\label{eq2}
\end{equation}
\begin{equation}
\mbox{\rm where}\ P_{\rm c}\Psi_L(\bm{R})=X(R)\Psi_L(-\bm{R})=
 (-1)^L X(R)\Psi_L(\bm{R}). \label{eq3}
\end{equation} 
The function $X(R)$ originates, in general, from both the transfer form factor and 
spectroscopic factor of the valence nucleon or cluster, and is radial dependent 
and complex \cite{Ful72,Fra80}. The phase $(-1)^{2I_{\rm c}}$ in Eq.~(\ref{eq2}) 
is implied by the spin statistics of the dinuclear wave function. In the \oc case, 
the two $^{12}$C cores are spinless and this phase can be dropped. Then, 
\begin{equation}
 T=\langle \phi|V|[1+(-1)^L X(R)]\Psi_L(\bm{R})\rangle=
 \langle \phi|V[1+(-1)^L X(R)]|\Psi_L(\bm{R})\rangle. \label{eq4}
\end{equation}
Consequently, the formal expression of the \AA OP for the single-channel 
OM calculation is
\begin{equation}
 V_\text{OP}=[1+(-1)^L X(R)]V(\bm{R}), \label{eq5}
\end{equation} 
where the second term is the CEP. The potential $V_\text{OP}$ is, in fact, similar 
to the ones derived by Fuller and McVoy \cite{Ful72}, and Frahn and Hussein 
\cite{Fra80}. For a direct reaction process, the parity-dependent CEP is 
closely associated with the transfer process that favors the transfer of a small 
number of nucleons. The earlier microscopic studies \cite{Lem79,Bay77,Bay86,Aoki83} 
have also suggested a strong (Majorana) core exchange term for the \AA systems 
with small mass difference. From the consideration leading to Eq.~(\ref{eq5}), 
we have assumed in the present study a local OP that contains the parity-independent potential referred to as Wigner term ($V_{\rm W}$) and parity-dependent 
one as Majorana term ($V_{\rm M}$)
\begin{equation}
V_\text{OP}(R)=V_{\rm W}({R})+(-1)^L V_{\rm M}({R}). \label{eq6}
\end{equation}

It is obvious from the kinematic illustration of elastic transfer in Fig.~\ref{f1}
that the total elastic scattering amplitude can be written as a coherent sum of the 
amplitudes of both the direct elastic scattering and elastic transfer 
\cite{vOe75,Fra80} 
\begin{equation}
f(\theta)=f_{\rm ES}(\theta)+f_{\rm ET}(\pi-\theta). \label{eq7}
\end{equation}   
Given the standard expansion of the direct elastic scattering (ES) amplitude 
into the partial wave series  
\begin{equation}
f_{\rm ES}(\theta)=f_{\rm R}(\theta)+\dfrac{1}{2ik}\sum_L(2L+1)e^{2i\sigma_L}
[S^{\rm (ES)}_L-1]P_L\left(\cos\theta\right), \label{eq8}
\end{equation}
where $f_{\rm R}(\theta)$ and $\sigma_l$ are the Rutherford scattering 
amplitude and Coulomb phase shift, respectively \cite{Sat83}, the elastic 
transfer (ET) amplitude can be expressed in the same manner     
\begin{eqnarray}
f_{\rm ET}(\theta)&=&\dfrac{1}{2ik}\sum_L(2L+1)e^{2i\sigma_L}
S^{\rm (ET)}_L P_L\left(\cos(\pi-\theta)\right) \nonumber\\
&=& \dfrac{1}{2ik}\sum_L(2L+1)e^{2i\sigma_L}S^{\rm (ET)}_L 
(-1)^L P_L\left(\cos\theta\right). \label{eq9}
\end{eqnarray}
The total elastic amplitude is then obtained as 
\begin{equation}
f(\theta)=f_{\rm R}(\theta)+\dfrac{1}{2ik}\sum_L(2L+1)e^{2i\sigma_L}
(S_L-1)P_L\left(\cos\theta\right), \label{eq10}
\end{equation}
\begin{equation}
\mbox{\rm where}\ S_L=S^{\rm (ES)}_L+(-1)^L S^{\rm (ET)}_L. \label{eq11}
\end{equation}
We have thus obtained the total elastic amplitude (\ref{eq10}) in the same 
partial-wave expansion as Eq.~(\ref{eq8}), but with a parity-dependent contribution 
from elastic transfer added to that of elastic scattering. The interference 
between these two terms gives rise naturally to an oscillating elastic cross 
section at large angles, similar to the Mott oscillation observed in elastic 
scattering of two identical nuclei like \cc or \oo \cite{Kho16}. Relation 
(\ref{eq11}) is a simplified version of the formal expression derived by Frahn 
and Hussein \cite{Fra80} for elastic transfer, where the impact of the dynamic 
$L$-dependent coupling potential caused by elastic transfer was shown to be 
equivalent to that of a modified elastic $S$-matrix that contains a parity-dependent component. Therefore, the assumption (\ref{eq6}) for the local OP
to be derived from the IP inversion of the elastic $S$-matrix is well
founded. 

\begin{figure}[bht!]\vspace*{0cm}
	\includegraphics[width=0.88\textwidth]{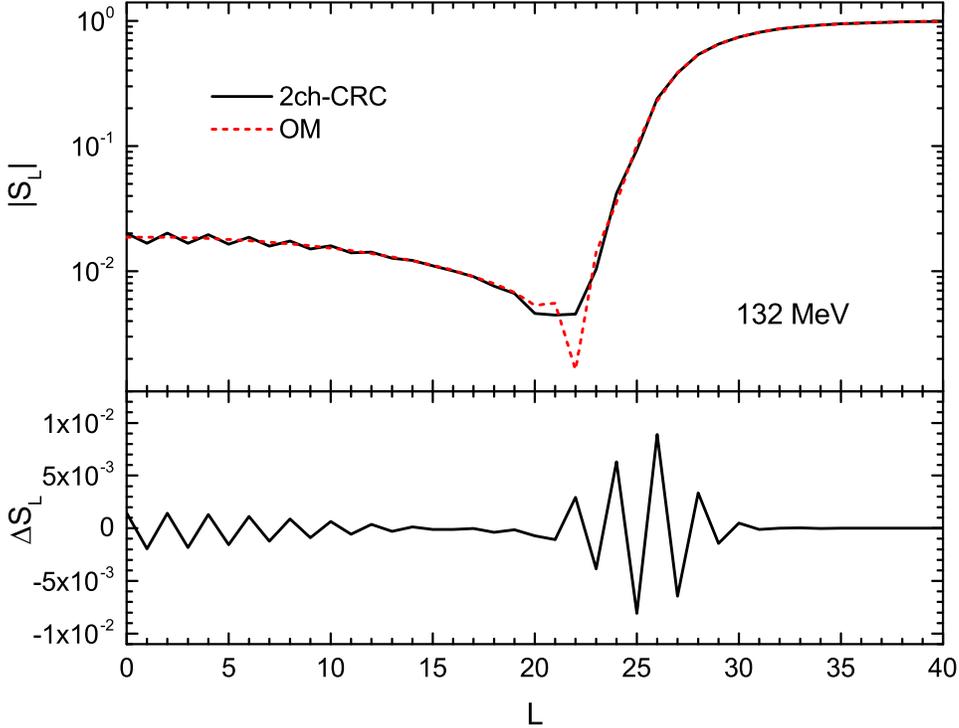}\vspace*{-0.5cm}
\caption{Modulus of the elastic $S$-matrix (upper panel) given by the single-channel 
OM (dashed line) and two-channel CRC (solid line) calculations of the elastic \oc 
scattering at $E_{\rm lab}=132$ MeV. The lower panel shows the difference 
$\Delta S_L=|S_L^\text{CRC}|-|S_L^\text{OM}|$.} \label{f2}
\end{figure}
\begin{figure}[bht!]\vspace*{0cm}
	\includegraphics[width=0.88\textwidth]{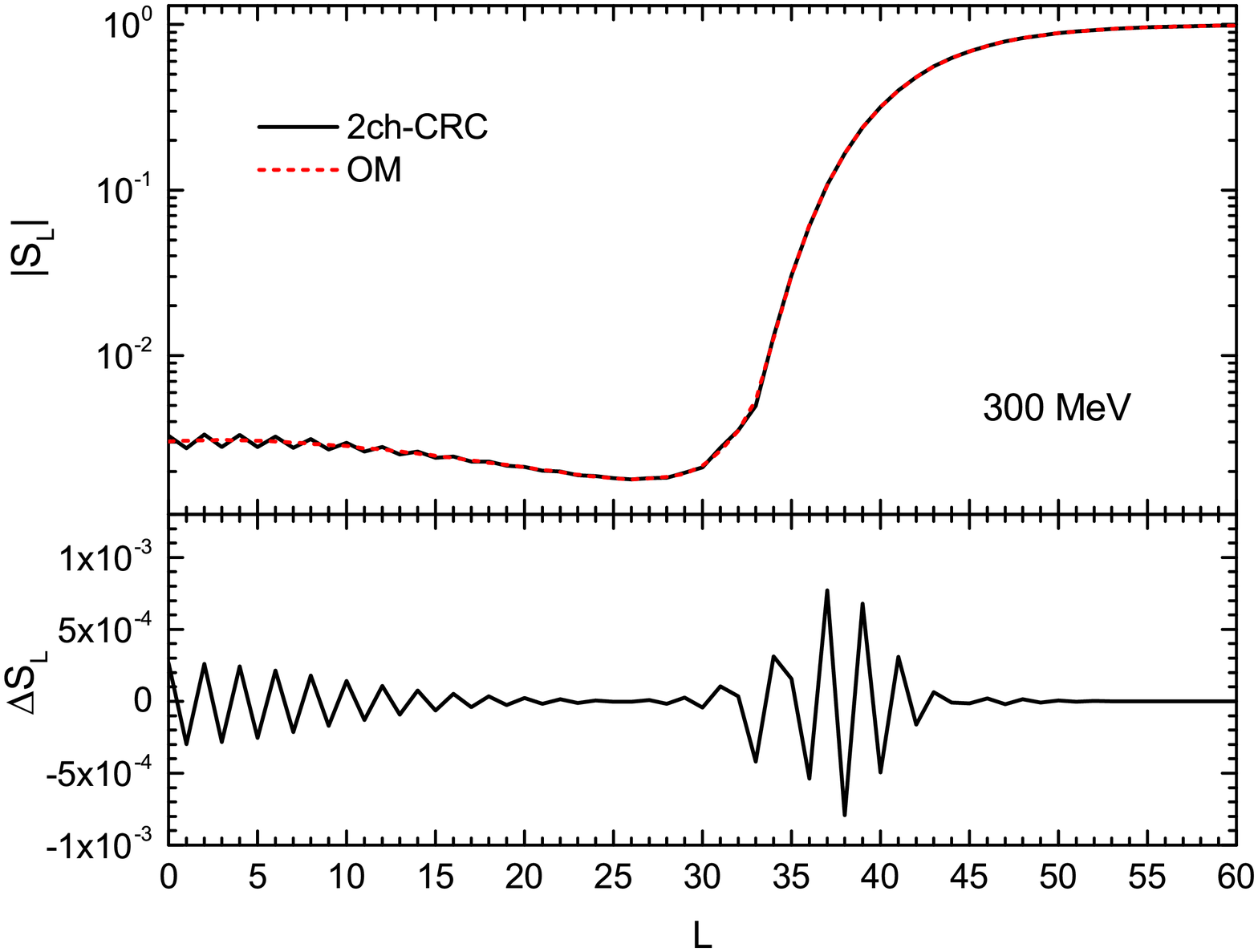}\vspace*{-0.5cm}
	\caption{The same as Fig.~\ref{f1} but for the elastic \oc scattering at 
	$E_{\rm lab}=300$ MeV.} \label{f3}
\end{figure}
For the illustration, we have plotted in Figs.~\ref{f2} and \ref{f3} the elastic 
$S$-matrix given by the recent two-channel CRC calculation of elastic \oc scattering 
at $E_{\rm lab}=132$ and 300 MeV \cite{Phu18} (with the direct elastic scattering 
and elastic $\alpha$ transfer channels explicitly taken into account) versus the $S$-matrix 
given by the single-channel OM calculation. One can see that with the elastic transfer 
channel taken into account, the total elastic $S$-matrix becomes parity-dependent 
as formally shown by relation (\ref{eq11}). This is also known as the odd-even staggering 
which is most pronounced around the grazing angular momenta ($L_g\approx 26$ and 37 for 
the 132 MeV and 300 MeV cases, respectively) as shown in lower panels of Figs.~\ref{f2} 
and \ref{f3}. Similar staggering occurs in the argument of the $S$-matrix which is 
related to the real potential. We note that the pattern of $\Delta S_L$ and a strong 
parity dependence of the elastic $S$-matrix near the grazing angular momenta are similar 
to the results obtained using the LCNO potential \cite{Ait93} for the $^{16}$O+$^{20}$Ne system. 
This range of grazing angular momenta corresponds to the forward-angle scattering caused 
by the \oc interaction at the surface, where the $\alpha$ transfer process was shown 
\cite{Phu18} to be dominant. The simple reason why the elastic $\alpha$ transfer 
(or the exchange of two identical cores) shows up in the enhanced oscillation of the 
elastic cross section at backward angles is that the elastic $\alpha$ transfer amplitude 
at $(\pi-\theta)$ is coherently added to the elastic scattering amplitude 
at $\theta$, as implied by the relation (\ref{eq7}).  

\begin{figure}[bht!]\vspace*{0cm}
	\includegraphics[width=0.8\textwidth]{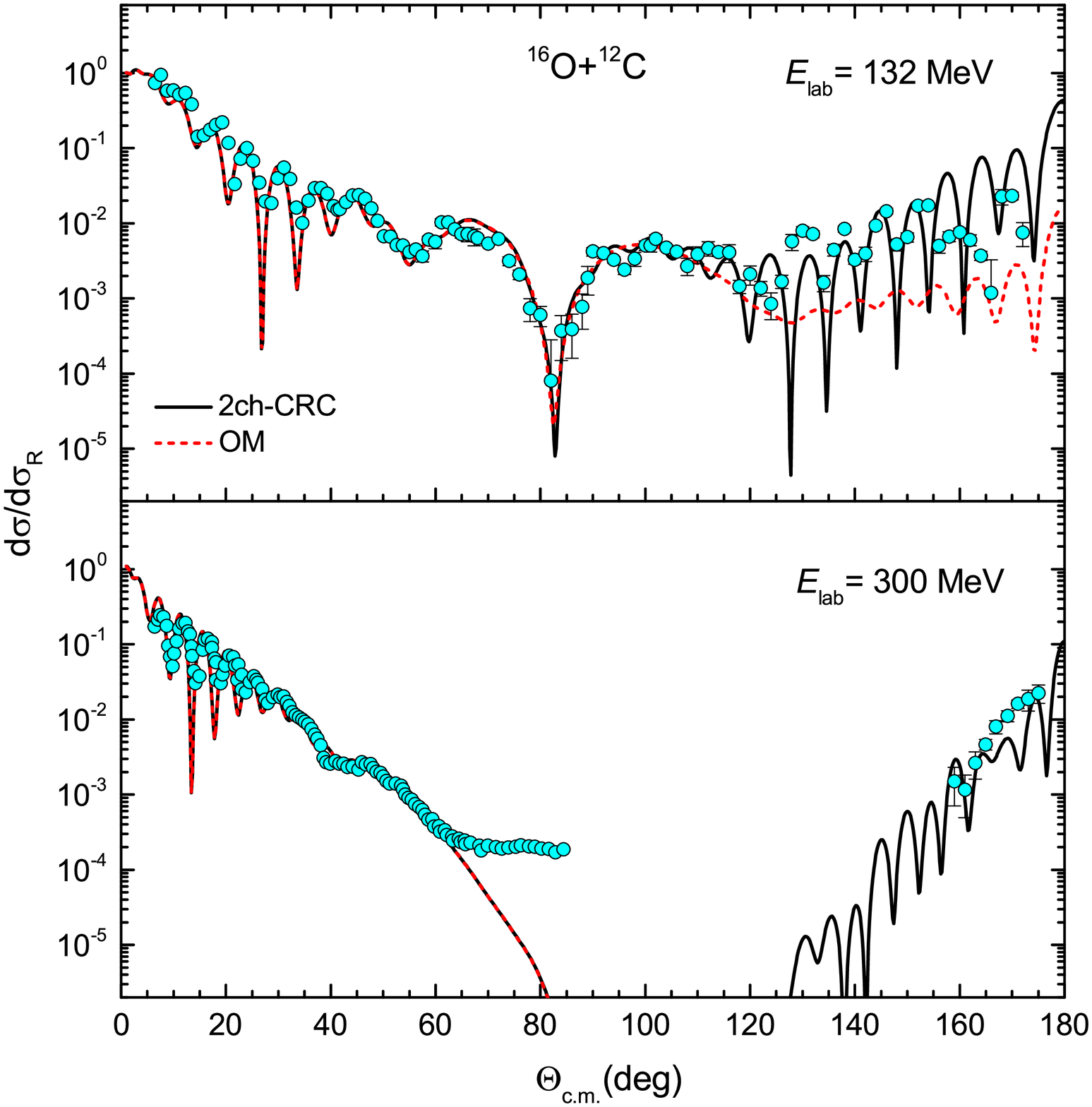}\vspace*{-1cm}
\caption{Elastic \oc scattering data measured at $E_\text{lab}=132$ MeV 
 \cite{Oglo00,Oglo98} and 300 MeV \cite{Bra01} in comparison with the results 
 given by the single-channel OM (dashed line) and two-channel CRC (solid line) 
 calculations \cite{Phu18}.} \label{f4}
\end{figure}
\section{CRC description of elastic alpha transfer}
\label{sec3}
In the present work we aim to derive explicitly the CEP generated by elastic 
$\alpha$ transfer based on the CRC description of the elastic \oc data measured 
at $E_\text{lab}=132$ MeV \cite{Oglo00,Oglo98} and 300 MeV \cite{Bra01}. The recent 
CRC study \cite{Phu18} has shown that elastic $\alpha$ transfer between the 
two identical $^{12}$C cores is the main physics origin of the enhanced oscillation 
of the elastic \oc cross section observed at backward angles at the two considered 
energies. The CRC results for the elastic \oc scattering including explicitly up to 
10 reaction channels for both the direct and indirect (multistep) $\alpha$ transfer 
account well for the measured data over the whole angular range, using the $\alpha$ 
spectroscopic factor $S_\alpha$ obtained from the large-scale shell model 
calculation by Volya and Tchuvilsky \cite{Volya15,Volya17}. 

\begin{figure}[bht!]\vspace*{0cm}
\includegraphics[width=0.8\textwidth]{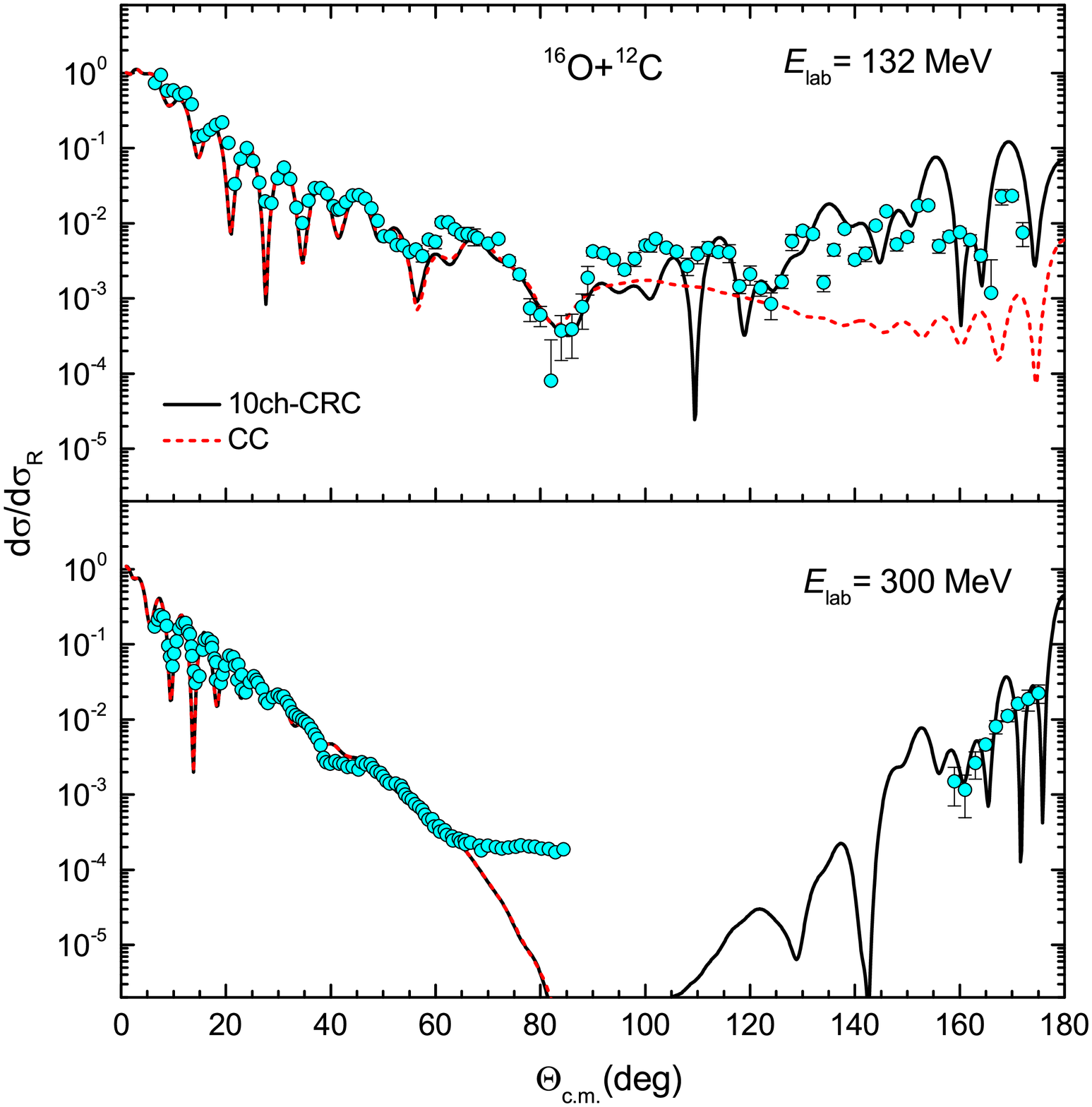}\vspace*{-1cm}
\caption{The same as Fig.~\ref{f4} but in comparison with the results given by the 
 coupled channel (dashed line) and ten-channel CRC (solid line) calculations 
 \cite{Phu18}.} \label{f5}
\end{figure}
We briefly discuss here the two coupling schemes for elastic $\alpha$ transfer used 
in our recent CRC analysis \cite{Phu18} of elastic \oc scattering using the code FRESCO 
\cite{Tho88}. The first scenario is the two-channel CRC calculation that includes 
only the true elastic scattering and direct elastic $\alpha$ transfer as considered in 
Eqs.~(\ref{eq7})-(\ref{eq11}). From results shown in Fig.~\ref{f4} one can see very
clearly the contribution of the elastic $\alpha$ transfer or core-core exchange 
showing up at backward angles. Such an approximation requires the minimum model space 
in the CRC calculation and explicitly generates the direct (one-step) exchange of the 
two identical $^{12}$C cores in elastic \oc scattering. Although the best-fit $\alpha$ 
spectroscopic factor obtained in the two-channel CRC analysis \cite{Phu18} is larger 
than that predicted by the structure studies, the strong effect of elastic $\alpha$ 
transfer revealed in this calculation, especially at the energy of 300 MeV, provides 
an important test ground for our approach to determine explicitly the CEP in 
the local OP (\ref{eq6}) for the one-channel OM description of these data.   

The second scenario is the ten-channel CRC description where the elastic scattering 
channel is coupled with the inelastic scattering channels for the $2^+_1$ (4.44 MeV) 
state of $^{12}$C, and $0^+_2$ (6.05 MeV), $3^-_1$ (6.13 MeV), and $2^+_1$ (6.92 MeV) 
states of $^{16}$O, and the direct and indirect $\alpha$ transfer channels through 
the considered excited states. The inelastic scattering form factors for the CRC 
calculation were calculated in the generalized double-folding model \cite{KhoS00} 
using the CDM3Y3 density dependent interaction \cite{Kho16}, and nuclear transition 
densities from the Resonating Group Method (RGM) by Kamimura \cite{Kam81} for $^{12}$C, 
and Orthogonality Condition Model by Okabe \cite{Okabe} for $^{16}$O. As a result, 
the model space of the ten-channel CRC configuration is quite large, and the total 
elastic cross section (see Fig.~\ref{f5}) is not a simple interference pattern 
(\ref{eq7}) of the two amplitudes but a superposition of all direct and indirect 
scattering and transfer amplitudes under consideration. The spurious deep minima 
between $120^\circ$ and $180^\circ$ in Fig.~\ref{f4} have been eliminated (see Fig.~\ref{f5})
through an interference of a large number of direct and indirect transfer amplitudes. 
With the measured elastic data well reproduced by the CRC calculation using the $\alpha$ 
spectroscopic factors given by the large-scale shell model calculation \cite{Volya15,Volya17}, 
the ten-channel CRC results shown in Fig.~\ref{f5} are deemed to be more realistic. 
Due to very small $\alpha$ spectroscopic factors predicted for the unbound excited states 
of $^{12}$C and $^{16}$O \cite{Volya15,Volya17}, the breakup effect to the ALAS should be
negligible, and we did not include the breakup channel into the model space
of the CRC calculation of elastic \oc scattering at low energies \cite{Phu18}. 

\section{Optical potential inverted from the elastic $S$-matrix}
\label{sec4}
To determine the equivalent CEP generated by elastic $\alpha$ transfer or core-core 
exchange, we have performed the inversion of the elastic scattering $S$-matrix given by the 
CRC calculation to a local, equivalent OP (\ref{eq6}). The CEP is obtained as the 
Majorana potential $V_{\rm M}(R)$ by substracting the Wigner potential $V_{\rm W}(R)$ from the 
inverted OP. The iterative perturbative (IP) inversion procedure \cite{IP1,IP2,IP3,IP4} 
implemented in the code IMAGO \cite{IMA} delivers a local complex OP that reproduces 
with very high precision the elastic scattering $S$-matrix given by the CRC calculation 
(referred to as the target $S$-matrix, $S^{\rm t}_L$). The accuracy of the inversion 
procedure is given by the quantity $\sigma$ defined as
\begin{equation}
\sigma^2=\sum_{L}\left|S^{\rm t}_L-S^{\rm i}_L\right|^2,
\end{equation} 
where $S^{\rm i}_L$ is the $S$-matrix for the potential found by the inversion process. 
The IP procedure can yield separate potentials for the even-$L$ partial waves and 
the odd-$L$ partial waves, $V_{\rm even}=V_{\rm W} + V_{\rm M}$ and 
$V_{\rm odd}=V_{\rm W} - V_{\rm M}$, with $V_{\rm W}(R)$ and $V_{\rm M}(R)$ are, 
respectively, the Wigner and Majorana components defined in Eq.~(\ref{eq6}).  

The inversion procedure begins the iterative process with a starting reference potential 
(SRP), which is usually the OP used in the original OM or coupled channel calculation. 
It has been found that inversion with the IP method leads to potentials that are generally 
independent of the SRP \cite{IP3,IP4}, as can be tested in particular cases. The IP inversion 
method was used earlier to investigate the parity dependence of the \AA OP for some core-identical 
systems like $^{3}$He/$t$+$\alpha$ \cite{Mac95} or $^{16}$O+$^{20}$Ne \cite{Ait93} at low 
energies. These studies have used, however, the $S$-matrices given by the models 
that are quite different from the CRC formalism. While $S^{\rm t}_L$ used in Ref.~\cite{Mac95} 
was taken from the RGM calculation, the one used in Ref.~\cite{Ait93} was given 
by the LCNO method that already includes a phenomenological parity-dependent 
potential into the real OP. The present work is the first attempt to determine the CEP 
for a light HI system at higher energies ($E>5$ MeV/nucleon) exclusively from the coupling 
between the elastic scattering channel and different inelastic scattering- and transfer 
reaction channels, by inverting the elastic $S$-matrix given by the CRC 
calculation of elastic \oc scattering.  

\begin{figure}[bht!]\vspace*{0cm}
	\includegraphics[width=0.8\textwidth]{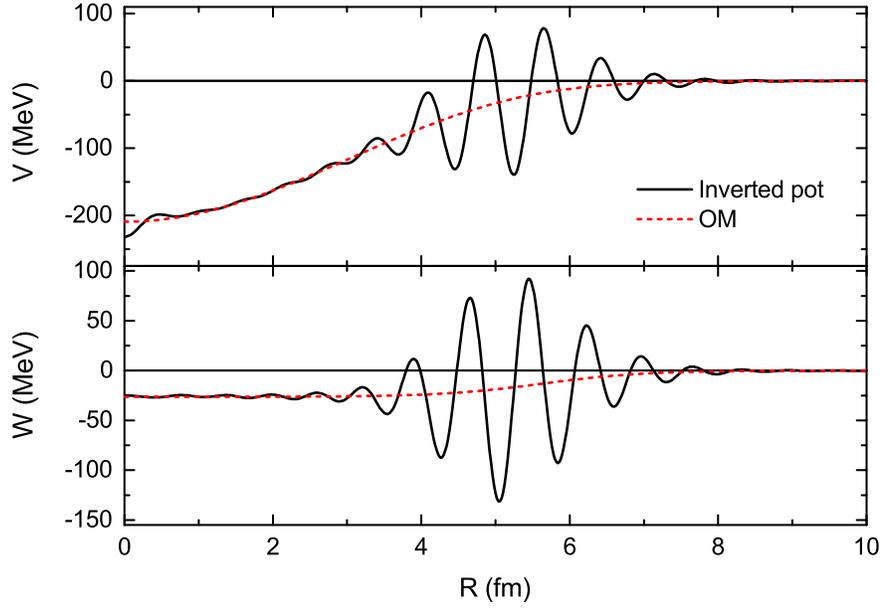}\vspace*{-0.5cm}
	\caption{OP assumed to contain only the Wigner term (solid line) inverted 
	from the $S$-matrix given by the two-channel CRC calculation of elastic \oc 
	scattering at $E_{\rm lab}=300$ MeV. The original OP or SRP is shown as dashed 
	line.} \label{f6}
\end{figure}
\begin{figure}[bht!]\vspace*{0cm}
	\includegraphics[width=0.8\textwidth]{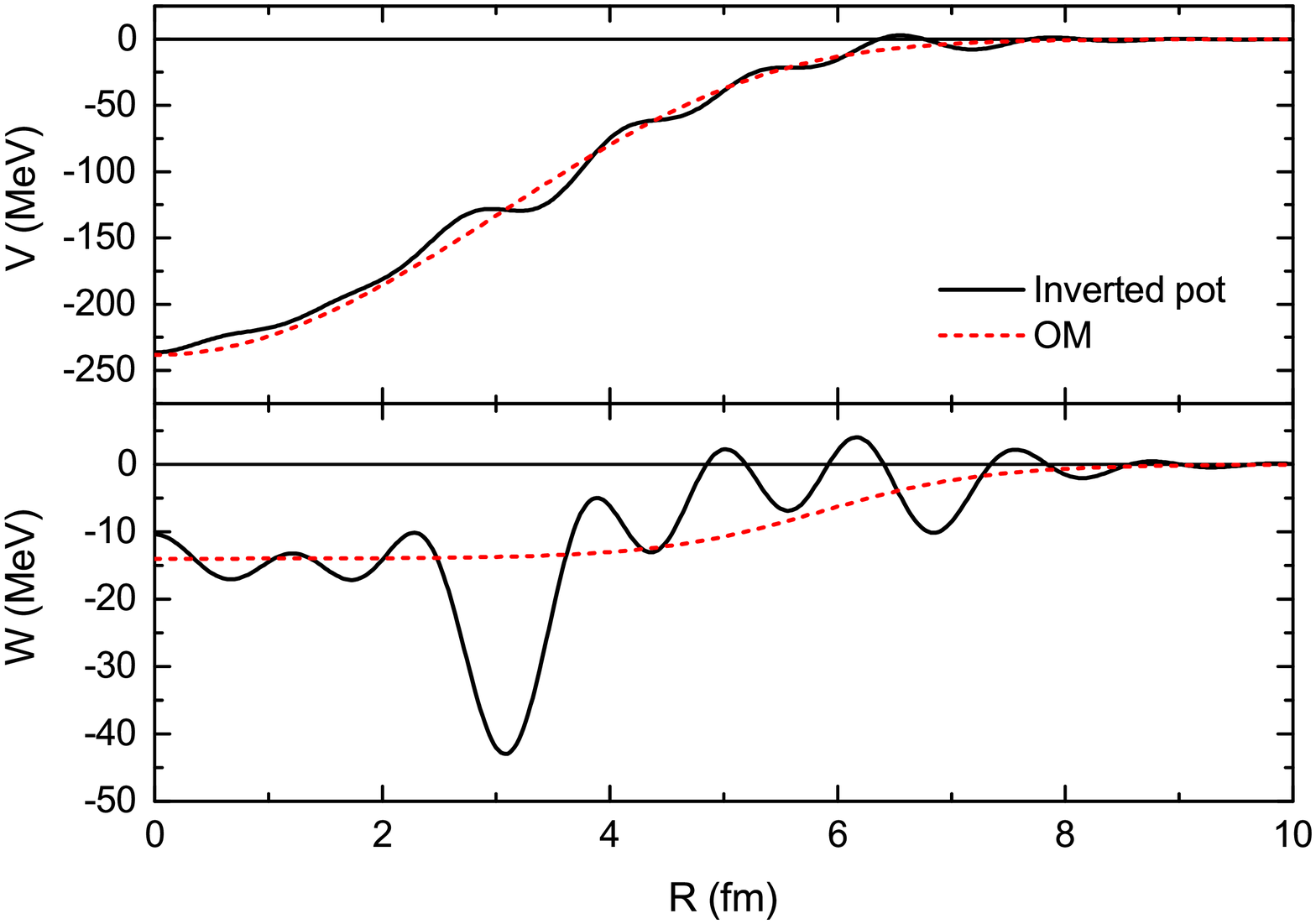}\vspace*{-0.5cm}
	\caption{The same as Fig.~\ref{f6} but for $E_{\rm lab}=132$ MeV.} \label{f7}
\end{figure}
\begin{figure}[bht!]\vspace*{-1cm}
	\includegraphics[width=0.8\textwidth]{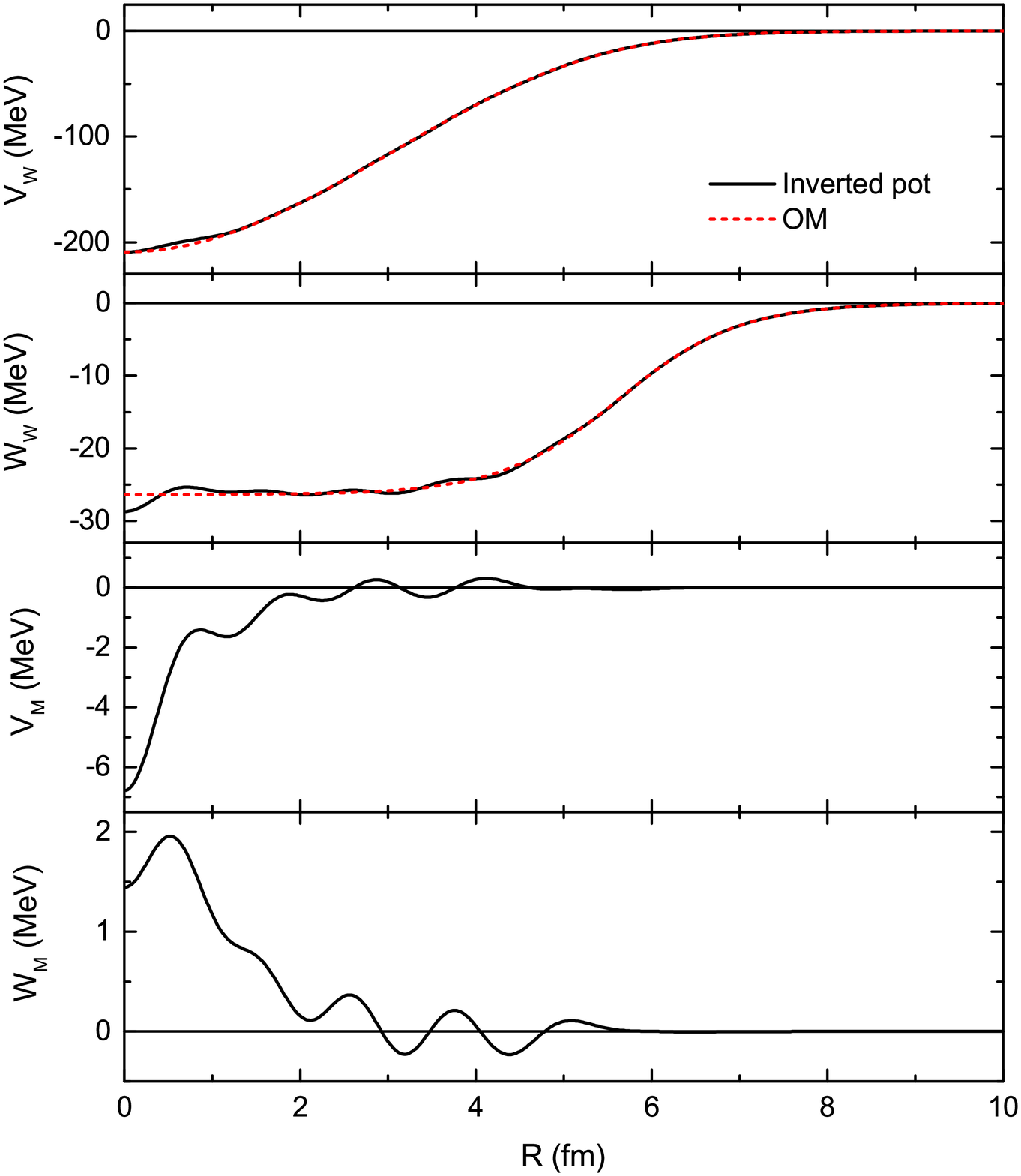}\vspace*{-1.5cm}
	\caption{OP assumed to contain both the Wigner and Majorana terms (solid line), 
	inverted from the $S$-matrix given by the two-channel CRC calculation of elastic 
	\oc scattering at $E_{\rm lab}=300$ MeV. The original OP or SRP is shown as dashed 
	line.} \label{f8}
\end{figure}
\begin{figure}[bht!]\vspace*{-1cm}
	\includegraphics[width=0.8\textwidth]{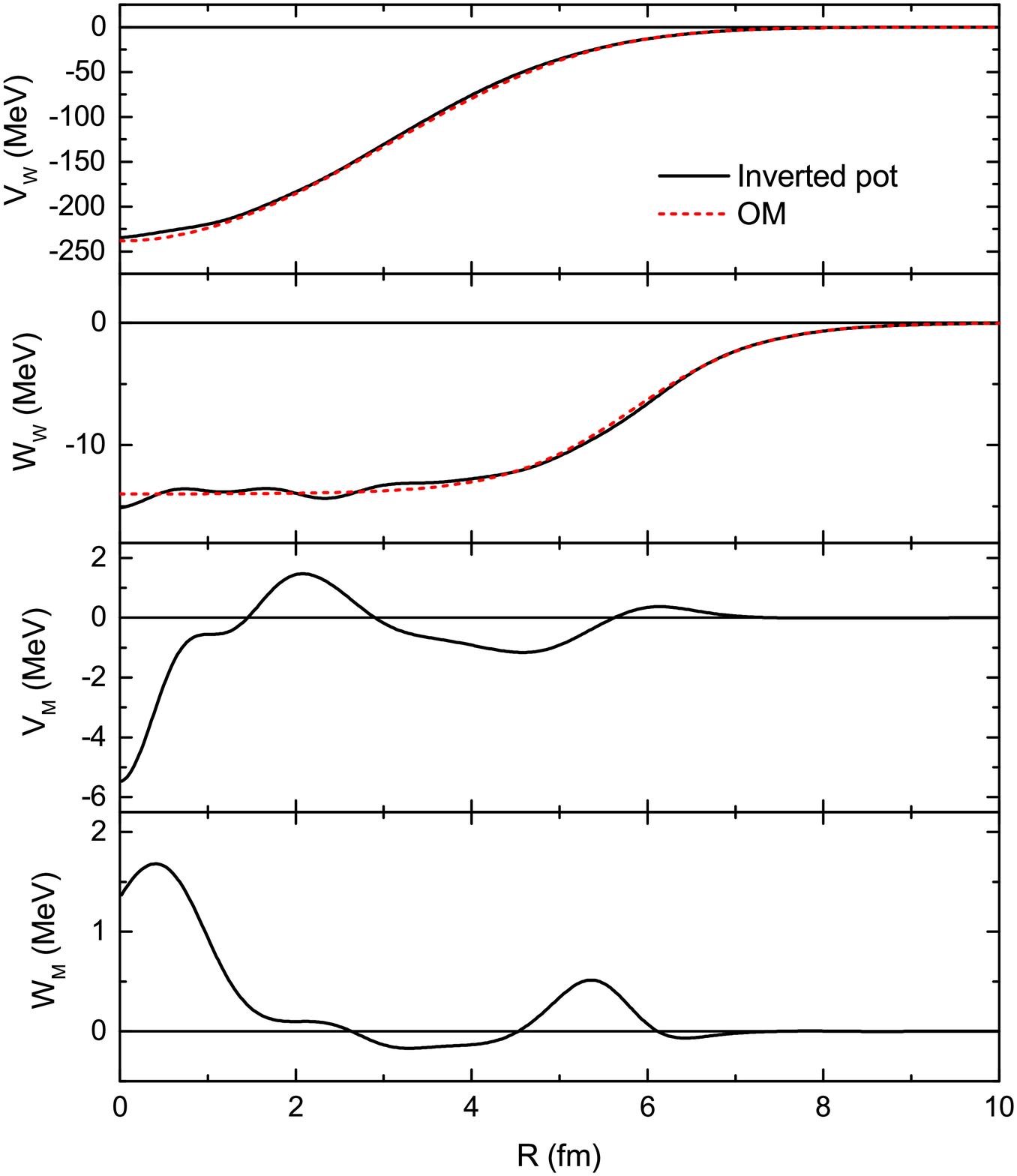}\vspace*{-1.5cm}
	\caption{The same as Fig.~\ref{f8} but for $E_{\rm lab}=132$ MeV.} \label{f9}
\end{figure}    
The reliability of the IP inversion method was tested first with $S^{\rm t}_L$ given 
by the single-channel OM calculation. In this case, the inverted OP is 
almost identical with the original OP and gives the elastic cross section 
that is graphically indistinguishable from that given by the OM calculation
(dashed line in Fig.~\ref{f4}). We discuss now the inversion results obtained 
with the target $S$-matrix given by the two-channel CRC calculation that 
includes only the true elastic scattering and direct elastic $\alpha$ transfer 
(solid line in Fig.~\ref{f4}). The necessity of a parity-dependent term in the total 
OP can be well illustrated by imposing a shape of the inverted OP in Eq.~(\ref{eq6}) that 
contains only the Wigner term. From the results shown in Figs.~\ref{f6} and \ref{f7}, 
one can see that both the real and imaginary parts of the inverted OP (obtained with 
a high precision of $\sigma=2.3\times10^{-3}$) are strongly undulatory at the surface 
($R\approx 3\sim 6$ fm), especially, at the energy of 300 MeV. This indicates clearly 
to the lack of a parity-dependent term in the total OP \cite{Coo90}, which is 
expected to peak in the surface region where the $\alpha$ transfer is dominant 
\cite{Phu18}. At 300 MeV, the distinctive ``V-shape'' cross section shown in lower 
panel of Fig.~\ref{f4} and a strongly localized oscillation of the inverted OP suggest 
that the parity dependence of the OP caused by the elastic $\alpha$ transfer or 
core-core exchange is more pronounced at this energy, where the data points at the 
most backward angles are entirely due to the $\alpha$ transfer and cannot be 
reproduced by a single-channel OM calculation \cite{Phu18}. At lower energies, 
the elastic $\alpha$ transfer and elastic scattering amplitudes are mixed at medium and 
large angles, and the enhanced oscillation of the elastic cross section can be reproduced 
in the single-channel OM calculation using a very small diffuseness of the imaginary 
Woods-Saxon (WS) potential \cite{Nico00}. Fig.~\ref{f7} shows that at the lower energy 
of 132 MeV the inverted OP is undulatory like that obtained in Ref.~\cite{Mac16} 
for the \oc system at $E_{\rm lab}=116$ MeV, where $S^{\rm t}_L$ was given by the OP 
with quite a small diffuseness of the imaginary WS potential. 

In order to investigate the parity dependence of the OP resulting from the core exchange 
process, we assume the shape of the local OP to contain both the Wigner and Majorana 
terms as in Eq.~(\ref{eq6}). The IP inversion with this prescription gives the inverted 
OP shown in Fig.~\ref{f8} ($\sigma=3.3\times 10^{-4}$) and Fig.~\ref{f9} ($\sigma=1.7\times 10^{-3}$) 
for $E_\text{lab}=300$ and 132 MeV, respectively. The inclusion of the \emph{parity-dependent} 
Majorana term into the OP significantly improves the accuracy of the inversion procedure, 
and the elastic $S$-matrix and scattering cross section given by the inverted OP in the 
single-channel OM calculation are nearly identical to those given by the two-channel 
CRC calculation. This similarity between the bare potential and the Wigner term of the inverted 
one is the direct evidence of the parity dependence of the CEP. One can see that both the real 
and imaginary parts of the Wigner potential become quite smooth when a complex Majorana term 
is included to account for the elastic $\alpha$ transfer. This means that the core exchange 
contribution to the elastic \oc cross section is well accounted for by the parity-dependent 
(Majorana) component of the OP. 

The results shown in Figs.~\ref{f8} and \ref{f9} are the direct representation of a complex 
parity-dependent CEP in the OP inverted for the single-channel OM calculation. The inverted 
real and imaginary Wigner parts are almost the same as those of the original OP. The elastic 
scattering cross section given by the single-channel OM calculation using the inverted OP  
is graphically indistinguishable from that given by the two-channel CRC calculation
(solid line in Fig.~\ref{f4}). We found a small oscillation of the Majorana potential
which should be associated with the elastic transfer form factor that by itself is 
undulatory \cite{vOe70,Ful72,Fra80}. The range of the Majorana term was found to be
slightly shorter than that of the Wigner term. The short range and weak strength of the CEP 
are expected features as suggested by the earlier microscopic study \cite{Bay86} and 
systematic parity-dependent analysis \cite{Fer90}. 

It is not surprising that the Majorana terms have their largest magnitude at small dinuclear
distances, as found in the earlier RGM calculations where the multi-nucleon exchange is 
included explicitly in the nucleon-nucleus \cite{Coo95} and nucleus-nucleus scattering 
\cite{Mac95,Aoki83}. It is reasonable that the parity-dependent term of the OP has the same 
behavior when the core exchange process is included explicitly in the CRC calculation. 
We have also done a test of cutting off the strength of the Majorana terms at small radii and 
found that the elastic \oc scattering cross section at the considered energies is sensitive 
to the Majorana potential mainly in the sub-surface region, with $R\gtrsim 3$ fm.    

\begin{figure}[bht!]\vspace*{-1cm}
	\includegraphics[width=0.8\textwidth]{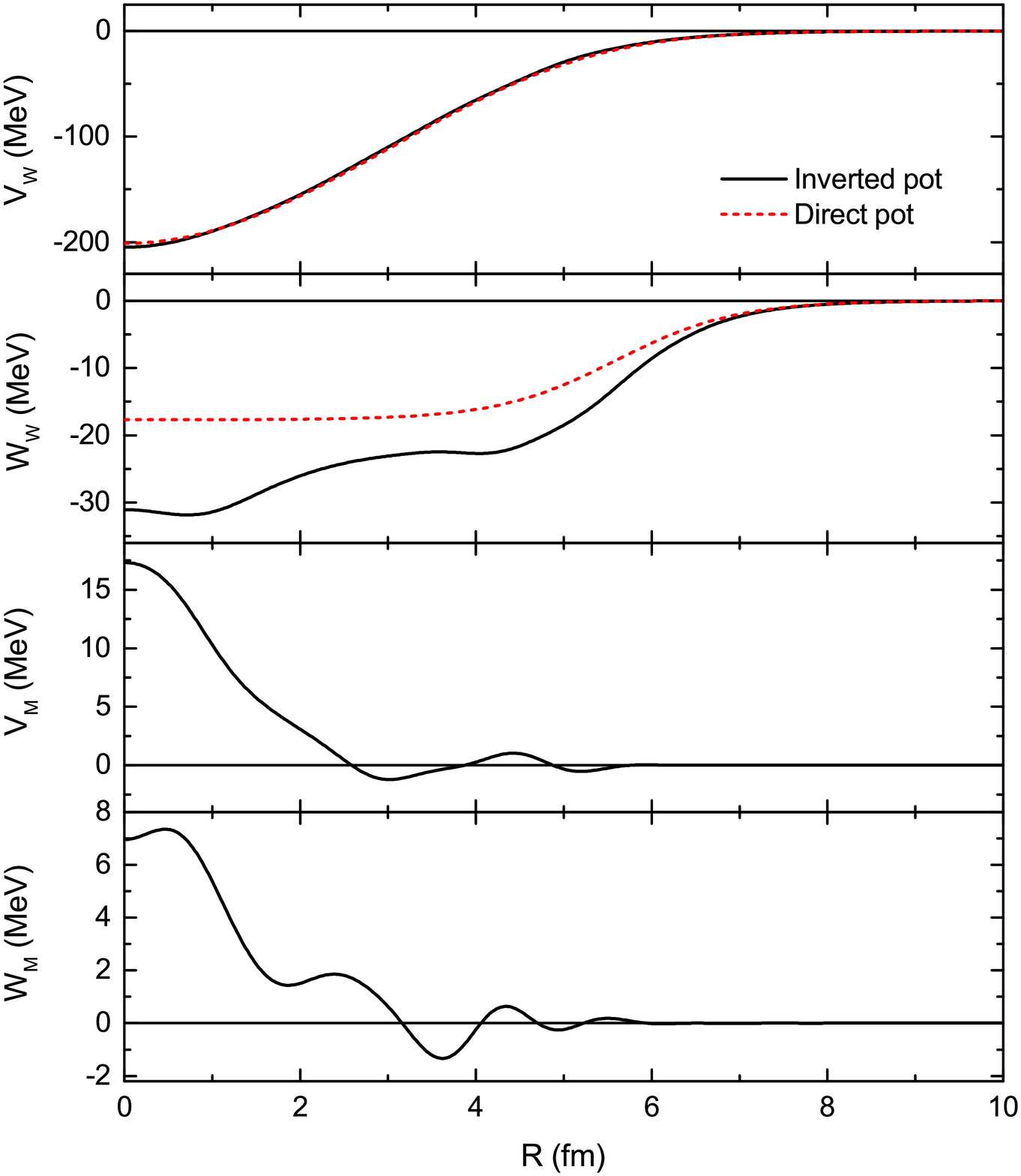}\vspace*{-1.5cm}
	\caption{OP assumed to contain both the Wigner and Majorana terms (solid line), 
	inverted from the $S$-matrix given by the ten-channel CRC calculation of elastic 
	\oc scattering at $E_{\rm lab}=300$ MeV. The original OP or SRP is shown as dashed 
	line.} \label{f10}
\end{figure}
\begin{figure}[bht!]\vspace*{1cm}
	\includegraphics[width=0.8\textwidth]{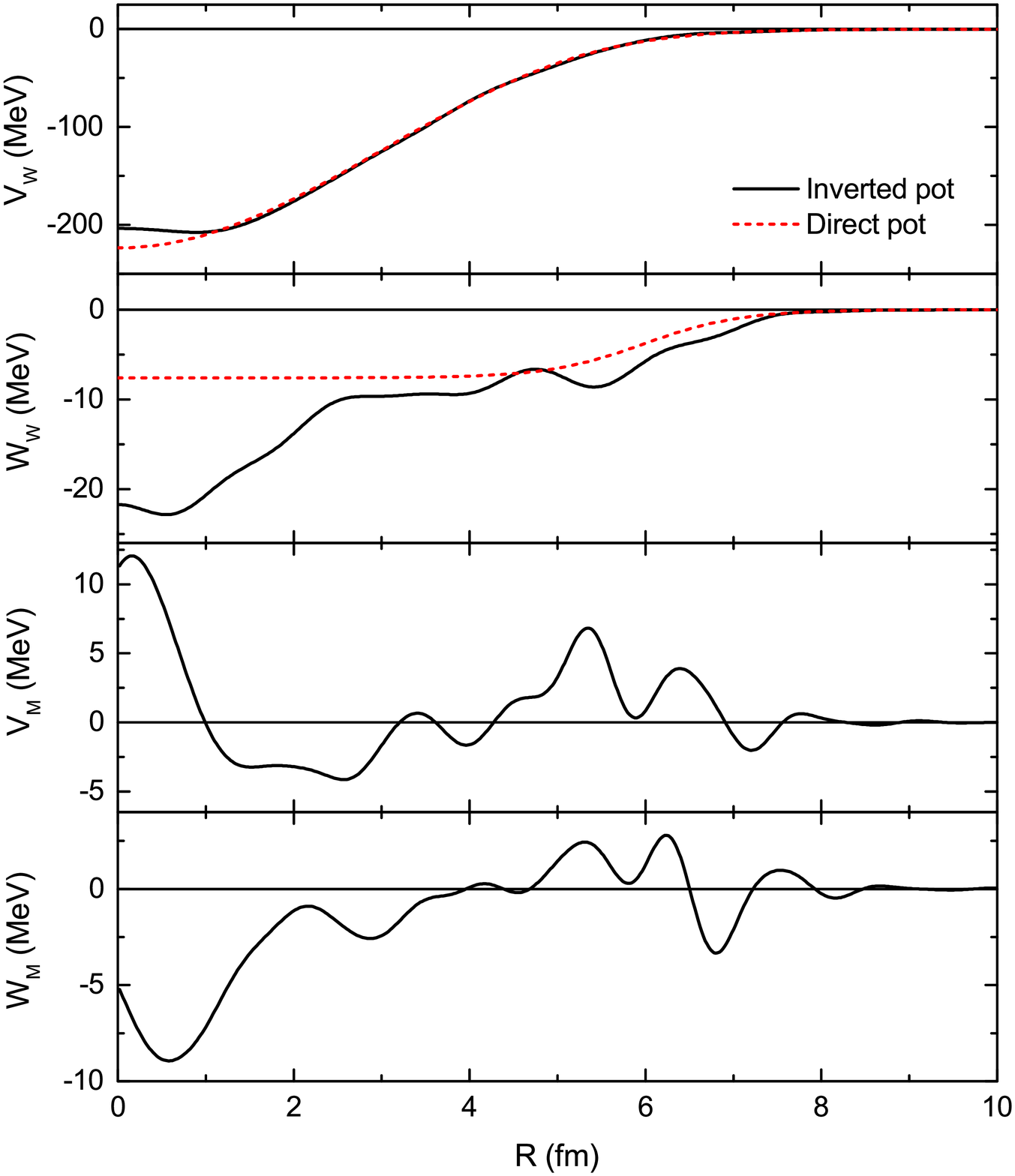}\vspace*{-1.5cm}
	\caption{The same as Fig.~\ref{f10} but for $E_{\rm lab}=132$ MeV.} \label{f11}
\end{figure}
The two-channel CRC calculation reproduces nicely the measured elastic data, but it 
requires an ``effective'' $S_\alpha$ that is much larger than that predicted by the 
structure studies. The main reason is that the two-channel CRC model space does not 
explicitly take into account the excitation of the two colliding nuclei. To have 
a more realistic estimate for the CEP given by the multistep process through 
different inelastic scattering and transfer channels, we have further performed 
the IP inversion of the $S$-matrix given by the ten-channel CRC calculation 
\cite{Phu18} (see Fig.~\ref{f5}). The inversion results are presented in 
Fig.~\ref{f10} ($\sigma=1.5\times 10^{-5}$) and Fig.~\ref{f11} 
($\sigma=1.4\times 10^{-4}$) for $E_\text{lab}=300$ and 132 MeV, respectively. 
Like the results obtained with the $S$-matrix given by the two-channel CRC calculation, 
the Wigner potential is quite smooth, with its real part being close to that of the 
original OP. On the other hand, the imaginary Wigner potential is deeper in the center
compared with the original OP. Such a difference in the imaginary OP is due to the dynamic 
polarization potential (DPP) arising from the coupling to the inelastic scattering channels. 
This coupling also increases the strength of the Majorana term and makes its structure more 
complicated compared to that obtained with the $S$-matrix given by the two-channel 
CRC calculation, especially, at the lower energy of 132 MeV (see Fig.~\ref{f11}). 

This work also shows how the inclusion of an explicit parity dependence makes it possible 
to identify the contribution of the collective excitations to the OP. The inversion of the 
$S$-matrix to the OP for the one-channel ON calculation has been used \cite{IP1,IP4} to determine
the contribution of inelastic channels or reaction channels to the nuclear OP. The core 
exchange process would make it impossible without allowing explicit parity dependence 
in the inversion. This is evident from the strong undularity seen in Figs.~\ref{f6} and \ref{f7}. 
The present work shows that the inclusion of a parity dependence into the inverted OP 
clearly reveals the contribution of the collective excitations to the OP (i.e. the DPP).
For the Wigner term in Fig.~\ref{f10}, the difference between the solid and dashed lines 
is a direct measure of the DPP due to the coupling to the inelastic scattering channels. 
Such a coupling leads to a very large absorptive term in the DPP, an increase of more than 50 
\% percent of the imaginary term at the origin. The contribution to the real part is small 
on the scale of the figure, but consistently repulsive by a few percent, except at the origin 
where it is slightly attractive. By contrast, Fig.~\ref{f8} shows that without the inelastic 
coupling there is effectively no enhanced absorption. Figure \ref{f11} shows that similar 
conclusions can be drawn at 132 MeV. Thus, our results show that the inclusion 
of the parity dependence enables the determination of DPP by the inversion 
in the presence of strong core exchange effects.  

\section{Summary}
The elastic $\alpha$ transfer or core exchange process in elastic \oc scattering
was shown to result in a parity-dependent CEP in the effective OP that gives
(in a single-channel OM calculation) the same description of elastic data 
over the whole angular range as that given by the CRC calculation that takes 
into account the core exchange explicitly. The high-precision IP inversion 
of the $S$-matrix given by the multichannel CRC calculation of elastic \oc 
scattering \cite{Phu18} gives readily a complex, parity-dependent Majorana 
potential that accounts for the core exchange in the \oc system. From a simple 
analytical derivation of the core exchange process, the parity dependence found 
in the OP inverted for a core-identical system is naturally explained. 
The inclusion of an explicit parity dependence also makes it possible to determine 
the DPP caused by the coupling to the collective excitations in the presence
of elastic $\alpha$ transfer. 

The complex structure of the obtained Majorana potential is likely associated 
with the properties of the elastic transfer or core exchange process, such as 
spectroscopic factors and transfer form factors. Therefore, the standard practice 
of using a simple prescription $[1+\alpha(-1)^L]V(R)$ in some phenomenological OM
studies cannot realistically represent the core exchange in elastic scattering 
of a core-identical system. 

We found that the $L$-dependence of the Majorana potential may be due partially 
also to the dynamic polarization of the OP by the coupling to different inelastic 
scattering channels. A more detailed study of the parity dependence of the \AA 
OP are, therefore, necessary and this will be the subject of our further research 
on this interesting and fundamental topic.    

\section*{Acknowledgments}
The present research has been supported, in part, by the National Foundation 
for Scientific and Technological Development of Vietnam 
(NAFOSTED Project No. 103.04-2017.317).

\end{document}